\pdfoutput = 1
\documentclass[sigconf]{acmart}

\usepackage{cite}
\usepackage{amsmath,amssymb,amsfonts}
\usepackage{algorithm, algorithmic}
\usepackage{graphicx}
\usepackage{textcomp}
\usepackage{xcolor}
\usepackage{pgfplots}
\usepackage{todonotes}
\usepgfplotslibrary{statistics}
\usepackage{subcaption}

\def\BibTeX{{\rm B\kern-.05em{\sc i\kern-.025em b}\kern-.08em
    T\kern-.1667em\lower.7ex\hbox{E}\kern-.125emX}}
    
\usepackage[normalem]{ulem} 
\usepackage{xcolor}
\usepackage{ifthen}
\usepackage{amssymb}
\newboolean{showcomments}

\setboolean{showcomments}{true}

\ifthenelse{\boolean{showcomments}}
{\newcommand{\nbc}[3]{
 {\colorbox{#3}{\bfseries\sffamily\scriptsize\textcolor{white}{#1}}}
 {\textcolor{#3}{\sf\small$\blacktriangleright$\textit{#2}$\blacktriangleleft$}}
 }
}
{\newcommand{\nbc}[3]{}
 }

\copyrightyear{2020}
\acmYear{2020}
\setcopyright{acmcopyright}
\acmConference[MSR '20]{17th International Conference on Mining Software Repositories}{October 5--6, 2020}{Seoul, Republic of Korea}
\acmBooktitle{17th International Conference on Mining Software Repositories (MSR '20), October 5--6, 2020, Seoul, Republic of Korea}
\acmPrice{15.00}
\acmDOI{10.1145/3379597.3387512}
\acmISBN{978-1-4503-7517-7/20/05}

\begin{document}

\title{An Exploratory Study to Find Motives Behind Cross-platform Forks from Software Heritage Dataset
}

\author{Avijit Bhattacharjee}
\affiliation{
\institution{Univeristy of Saskatchewan}
}
\orcid{0000-0002-9468-8816}

\author{Sristy Sumana Nath}
\affiliation{
\institution{Univeristy of Saskatchewan}
}

\author{Shurui Zhou}
\affiliation{
\institution{Carnegie Mellon University}
}

\author{Debasish Chakroborti}
\affiliation{
\institution{Univeristy of Saskatchewan}
}

\author{Banani Roy}
\affiliation{
\institution{Univeristy of Saskatchewan}
}

\author{Chanchal K. Roy}
\affiliation{
\institution{Univeristy of Saskatchewan}
}

\author{Kevin Schneider}
\affiliation{
\institution{Univeristy of Saskatchewan}
}
\renewcommand{\shortauthors}{A. Bhattacharjee, et al.}

\begin{abstract}
The fork-based development mechanism provides the flexibility and the unified processes for software teams to collaborate easily in a distributed setting without too much coordination overhead. Currently, multiple social coding platforms support fork-based development, such as GitHub, GitLab, and Bitbucket. Although these different platforms virtually share the same features, they have different emphasis. As GitHub is the most popular platform and the corresponding data is publicly available, most of the current studies are focusing on GitHub hosted projects. However, we observed anecdote evidences that people are confused about choosing among these platforms, and some projects are migrating from one platform to another, and the reasons behind these activities remain unknown. With the advances of Software Heritage Graph Dataset (SWHGD), we have the opportunity to investigate the forking activities across platforms. In this paper, we conduct an exploratory study on 10 popular open-source projects to identify cross-platform forks and investigate the motivation behind. Preliminary result shows that cross-platform forks do exist. For the 10 subject systems in this study, we found 81,357 forks in total among which 179 forks are on GitLab. Based on our qualitative analysis, we found that most of the cross-platform forks that we identified are mirrors of the repositories on another platform, but we still find cases that were created due to preference of using certain functionalities (e.g. Continuous Integration (CI)) supported by different platforms. This study lays the foundation of future research directions, such as understanding the differences between platforms and supporting cross-platform collaboration. 


\end{abstract}
\maketitle


\section{Introduction}

Fork-based development allows developers to start development from an existing codebase while having the flexibility and independence to make changes~\citep{zhou2018identifying,gousios2014exploratory}. 
Prior work studied the fork-based development mechanism from different perspectives, such as the collaboration efficiencies of software teams using forks~\citep{zhou2019fork}, pull request management processes~\citep{gousios2014exploratory}, sustainability of open-source communities~\citep{steinmacher2018almost}, and different types of forks on GitHub~\citep{zhou2020forkChange}. 
However, most of these studies focus on GitHub. Although GitHub is the most popular platform that supports fork-based development, other reasonably popular platforms support the fork-based development mechanism as well, such as BitBucket and GitLab. BitBucket announced that till April 2019 they reached 10 million registered users and over 28 million repositories\footnote{https://en.wikipedia.org/wiki/Bitbucket}. Similarly, GitLab is used by more than 100,000 organizations and its open-source codebase is contributed by 2,988 developers\footnote{https://about.gitlab.com/company/}. These platforms have different functionality emphasis such as DevOps solution, self-managed hosting, automatic code change monitoring, and Continuous Integration/Continuous Deployment (CI/CD). 
Anecdotal evidences show that people are struggling about which platform to choose for software development, and we observed projects migrating from one platform to another, such as the project \emph{Vim}~\citep{vim} was hosted on GitHub, and then was moved to GitLab and was renamed as \emph{mg-vim}~\citep{mgvim} because of the CI feature of GitLab.

Prior work on understanding different types of forks defined two main types of active forks on GitHub~\citep{zhou2019fork}: \textbf{social fork}, which is created to contribute back to the main repository, and \textbf{hard fork}, which is aiming for supersede or replace the original project. With the advances of the SWHGD~\citep{pietri2019software,MSR20DC}, which brings  different social coding platforms data under single roof including \emph{GitHub}, \emph{GitLab}, \emph{Debian}, and \emph{PyPI},  we have the opportunity to study the forking activity among different social coding platforms. Specifically, we would like to understand why developers prefer one platform over another, what are the features and limitations of platforms which drive the migration of code, and see if there are space of improvement to current fork-based development mechanism.

In a nutshell, this paper conducts an exploratory empirical study using mixed methods (qualitative and quantitative) on ten popular projects to address the following two research questions:  

\begin{itemize}
    \item RQ1. How often do cross-platform forks happen?
    \item RQ2. What are the motives behind cross-platform forks?
\end{itemize}

The contributions of this paper are as follows. (1) We conducted the first-ever empirical study on cross-platform forks, (2) we propose an algorithm that automatically detects cross-platform forks, and (3) we provide a foundation of future research directions towards finding active cross-platform forks automatically and tracking their activities.

\section{RQ1: How often do cross-platform forks happen?}
We would like to understand the frequency of cross-platform forks. To achieve our goal, we designed a two-step quantitative study to automatically detect the cross-platform fork candidates. First, we tried to detect forks of a target origin from the SWHGD. Second, we filtered out the forks of different platforms to find cross-platform forks. From our preliminary analysis with 10 representative projects from GitHub, we found 179 cross-platform forks from our subject systems. 

\subsection{Subject Systems and Data Collection}
\paragraph{SWHGD}
The \emph{origin} table of the SWHGD~\citep{MSR20DC} contains URLs of the repositories crawled. After some time interval, URLs of the origin table are visited again to capture new updates. These visits to an origin URL is stored in the \emph{origin\_visit} table. The \emph{snapshot} table contains information about snapshots of an origin URL. Branches associated with each snapshot are stored in the \emph{snapshot\_branches} table. The \emph{snapshot\_branch} table stores the commits each branch of a snapshot points to. Finally, all commits of an origin URL are stored in the \emph{revision} table which contains information about individual commits (i.e. author, committer, date, and message).
\paragraph{Sampled dataset}
We randomly selected 10 projects on GitHub with a different number of forks using GHTorrent~\citep{Gousi13}. To not bias our analysis by practices applied by the largest or
by many small projects, we sampled 5 very frequently forked projects and 5 moderately forked projects, as shown in Table ~\ref{table:considered-repos}.
\begin{table} 
 \caption{10 subject systems with number of forks from the GitHub API, the SWHGD and time required to extract from SWHGD}
\centering
\resizebox{3.4in}{!}{
\begin{tabular}{l|l|l|l|l|l}
No & URL & Name & No. of Forks & No. of forks  & Time(Hr)\\
 & (https://github.com) &  & (From GitHub) & (From SWHGD)  & \\
\hline
1 & \url{/sloria/TextBlob}& TextBlob & 903 & 513 & 27min \\
2 & \url{/explosion/spaCy}& spaCy & 2700 & 756 & 45min \\
3 & \url{/flutter/flutter}& Flutter & 1080 & 483 & 8hr 7min \\
4 & \url{/vim/vim}& Vim & 2600 & 1339 & 12hr \\
5 & \url{/neovim/neovim}& Neovim & 2600 & 2284 & 5hr 15min \\
6 & \url{/bitcoin/bitcoin}& Bitcoin & 25000 & 10228 & 29hr 17min \\
7 & \url{/scikit-learn/scikit-learn} & scikit-learn & 19000 & 10551 & 10h 33min\\
8 & \url{/facebook/react-native} & ReactNative & 18800 & 11785 & 16hr 5min \\
9 & \url{/nodejs/node}& Node.js & 16000 & 15510 & 30hr 12min \\
10 & \url{/tensorflow/tensorflow} & TensorFlow & 79400 & 27908 & 21hr 8min \\
\end{tabular}
} 
\label{table:considered-repos}
\end{table}
\paragraph{Experiment setup}
We downloaded the full version of Software Heritage Dataset\footnote{https://annex.softwareheritage.org/public/dataset/graph/latest/sql/} in a VM with 24-core processor and 64GB RAM. The physical machine has dual Intel(R) Xeon(R) CPU E5-2680 v4 @ 2.40GHz CPUs, with a total of 28 cores (56 threads). We used a local PostgreSQL server to load the instance. During the indexing process, we omitted the unnecessary tables to save time. We used Python code to run SQL queries on the PostgreSQL instance using Sqlalchemy \citep{sqlalchemy} module. We uploaded the extracted data~\citep{cross_fork_dataset} and used Python program~\citep{cross_fork_software} to Zendoo. 




\begin{algorithm}
  \caption{SQL queries to retrieve forks of a target URL}
  \label{alg:the_alg}
  \begin{algorithmic}[1]
    \STATE \textbf{select} id \textbf{as} origin\_id \textbf{from} origin \textbf{where} url = :target\_url \label{op0}
    \STATE \textbf{select} snapshot\_branch.target \textbf{as} interval\_commits \textbf{from} origin\_visit, snapshot\_branches, snapshot\_branch \textbf{where} snapshot\_branch.object\_id = snapshot\_branches.branch\_id \textbf{and} snapshot\_branches.snapshot\_id = origin\_visit.snapshot\_id \textbf{and} origin = :origin \textbf{and} status = `full' \label{op1}
    \STATE \textbf{select distinct} id \textbf{as} child\_commits \textbf{from} revision\_history \textbf{where} parent\_id =  ANY(:interval\_commits) \label{op2}
    \STATE \textbf{select} url, snapshot\_branch.target \textbf{as} rev \textbf{from} origin, origin\_visit, snapshot\_branches, snapshot\_branch \textbf{where} origin.id = origin\_visit.origin \textbf{and} origin\_visit.snapshot\_id = snapshot\_branches.snapshot\_id \textbf{and} snapshot\_branches.branch\_id = snapshot\_branch.object\_id \textbf{and} snapshot\_branch.target  = ANY(:interval\_commits + :child\_commits)  \textbf{and} url != :target\_url  \label{op3}
   
  \end{algorithmic}
\end{algorithm}

\subsection{Research  Method}
 
 As SWHGD uses a deduplication technique to avoid storage of the same snapshot for multiple forks, all the forks including the original repository points to the same snapshot which allows us to detect forks over multiple platforms. 
To achieve generalization, SWHGD only stores project-specific data (commit, directory, release, snapshot, branch), and skipped platform dependent data such as issues, pull requests, forks, and others. 
Therefore, in order to detect forks, we need to rely on the comparison of commit history information and corresponding platform URL to find the cross-platform fork candidates. Specifically, we detect cross-platform forks by analyzing the following tables from the SWHGD: \emph{origin}, \emph{origin\_visit}, \emph{snapshot\_branches}, and \emph{snapshot\_branch}\footnote{We are aware of the method of combining both GHTorrent data and SWHGD data together to find forks, but there are inconsistencies that need more manual work to resolve, so in this study, we only focus on the SWHGD.}. 
 
\paragraph{Step 1: Finding forks from commit information}

 As shown in Algorithm \ref{alg:the_alg}, we detected forks of a repository from the SWHGD step by step. In (\ref{op0}), id of the target origin URL is retrieved. In (\ref{op1}), we extracted interval commits (last commit of each branch during a snapshot stored in \emph{snapshot\_branch} table) of an origin URL by joining \emph{origin\_visit}, \emph{snapshot\_branches}, \emph{snapshot\_branch} tables. (\ref{op2}) is to find child commits of the origin interval commits by searching all commits in \emph{revision\_history} table. Now, we need to map all the commits to their corresponding URL so that we can get the forks for our target URL. Before going to  (\ref{op3}), we merged the interval commits and child commits from (\ref{op1}), (\ref{op2}) into a target commit list. Finally, the query of (\ref{op3}) is used to map target commits with their corresponding URL by joining \emph{origin}, \emph{origin\_visit}, \emph{snapshot\_branches}, and \emph{snapshot\_branch} tables where commits are in target commit list and URL is different from our target URL. For more details of the approach, we would recommend the interested readers to look into our code made available via Zendoo~\citep{cross_fork_software}.
\begin{figure}
    \centering
    \begin{tikzpicture}
        \begin{axis}[
            stack plots=x,
            /tikz/xbar,
            ytick = {0, 1, 2, 3, 4, 5, 6, 7, 8, 9},
            yticklabels = {TextBlob, spaCy, Flutter, Vim, Neovim, Bitcoin, scikit-learn, ReactNative, Node.js, TensorFlow},
            legend style={at={(0.77,-0.10)}},
            yticklabel style={font=\small},
            xmin = 0,
            xmax = 100,
            ]
           
            \addplot coordinates {
                (56.81, 0) (28,1) (44.72 ,2) (51.5, 3) (87.84, 4) (40.91, 5) (55.53, 6) (62.68, 7) (96.93, 8) (35.14, 9)
            };
            
            \legend{
             Forks by our approach,
            };
        \end{axis}
        \end{tikzpicture}
    \caption{Percentage of forks retrieved by our approach}
    \label{fig:percentage_of_forks_retrieved}
\end{figure}
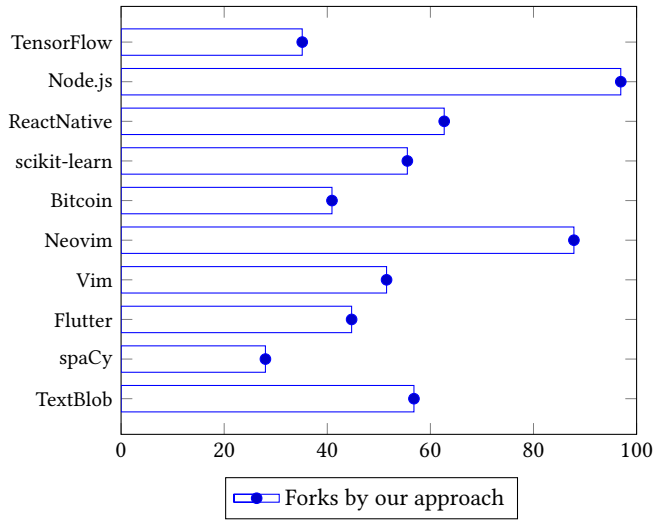
We only selected commits pointed by \emph{snapshot\_branch} table to find forks. The reason behind this choice is when we select all commits of an origin URL, the query to map commits to their URL takes a long time which hinders our approach to investigate motives behind cross-platform forks. As the commit to URL map query of (\ref{op3}) of Algorithm \ref{alg:the_alg} joins four large tables, the increase in number of commits increases the time required for the query. For example, when we tried to get forks for \emph{spaCy} project using all the commits, we have found 802 forks which is a little bit more compared to the numbers reported in Table \ref{table:considered-repos}. However, the time taken for the algorithm is 12 hours where considering interval commits from the \emph{snapshot\_branch} table takes 45 minutes to find 756 forks. 
\vspace{-2mm}
\paragraph{Step 2: Finding cross-platform forks}
In the second step, we used the forks extracted from the previous step to detect forks of a project on different platforms. While extracting forks, we used the commit information to detect whether a repository is a fork of a target origin. Hence, we expect to get forks from platforms different than GitHub, as our subject systems are from GitHub. Therefore, we filtered out the forks where URL has \emph{https://github.com} as a prefix of their URL. We reported the number of cross-platform forks in Figure \ref{fig:forks_from_gitlab}.
\subsection{Findings}
 In Table \ref{table:considered-repos}, we presented URL of the repositories, with the number of forks collected from GitHub and our approach. We also reported time taken by our program to search forks from \emph{\textbf{88,290,221}} repositories stored in the \emph{origin} table of the SWHGD. In Figure \ref{fig:percentage_of_forks_retrieved}, we plotted the percentage of forks we have been able to detect compared to the data from GitHub. We can see that for \emph{Node.js} and \emph{Neovim} our approach detected 87\% forks reported in GitHub. For \emph{TextBlob}, \emph{Vim}, \emph{scikit-learn}, and \emph{ReactNative} our approach detected approximately 56\%-63\% forks. For \emph{spaCy}, \emph{Vim}, \emph{Bitcoin} and \emph{TensorFlow} the percentage of forks is below 50\%. As we only used interval commits stored in the \emph{snapshot\_branch} table and their child commits from the \emph{revision\_history} table, the number of detected forks depends on how many forks points to the interval commits and their child commits.

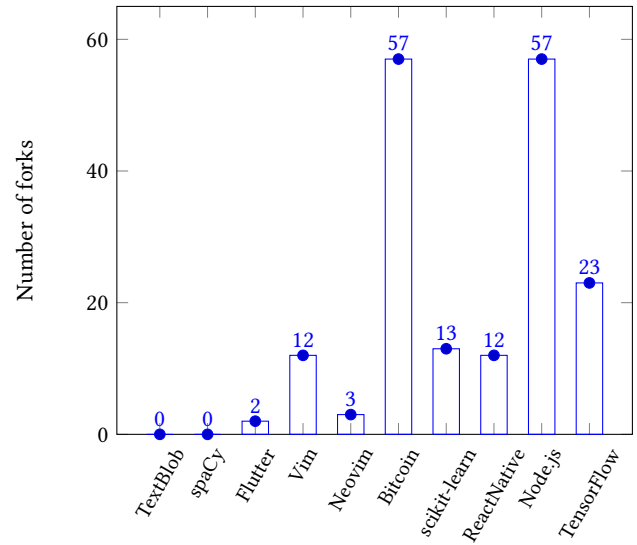
\begin{figure}
    \centering
    \begin{tikzpicture}
        \begin{axis}[
            stack plots=y,
            /tikz/ybar,
            xtick = {0, 1, 2, 3, 4, 5, 6, 7, 8, 9},
            xticklabels = {TextBlob, spaCy, Flutter, Vim, Neovim, Bitcoin, scikit-learn, ReactNative, Node.js, TensorFlow},
            legend style={at={(0.77,-0.10)}},
            xticklabel style={font=\small},
            ymin = 0,
            ymax = 65,
            xticklabel style={rotate=60},
            ylabel = Number of forks,
            nodes near coords,
            ]
           
            \addplot coordinates {
                (0, 0) (1, 0) (2,2) (3,12) (4, 3) (5, 57) (6,13) (7,12) (8,57) (9,23)
            };
            
        \end{axis}
        \end{tikzpicture}
    \caption{Forks detected from GitLab by our approach}
    \label{fig:forks_from_gitlab}
\end{figure}

Figure \ref{fig:forks_from_gitlab} plots the number of forks our approach extracted from GitLab platform. For \emph{Bitcoin} and \emph{Node.js} library, we found 57 forks from GitLab which is the highest among all our subject systems. \emph{TensorFlow} repository got the 2nd position among 10 subject systems by getting 23 forks from GitLab. \emph{ReactNative}, \emph{scikit-learn} and \emph{Vim} got 12-13 forks from GitLab. Whereas \emph{spaCy},  \emph{TextBlob} have no forks from GitLab. In the following section, we conducted a manual analysis to identify the motives behind the detected GitLab forks.



\section{RQ2. What are the motives behind cross-platform forks?}
In 2018, when Microsoft acquisition of GitHub took place, GitLab reported a spike of importing GitHub projects to their platform~\citep{githubDitched}. As many open-source developers believe that Microsoft is not open source friendly, the acquisition of GitHub might have motivated some of the developers to move their projects to GitLab. This is one of the motivations which caused cross-platform forks. Still, there might be other reasons for cross-platform forks. To better understand the motivation behind cross-platform forks, we conducted a qualitative study to find motives behind switching between platforms.
\subsection{Research  Method} 
To answer RQ2, two of the authors of this paper manually compared cross-platform forks with their origin repositories. For comparison between two repositories, title, description, commit history, and username of the fork owner are considered. To remove the subjectivity issue, two authors individually noted above-mentioned properties for studied 100 GitLab projects and their corresponding origin repositories. Later, a third author analyzed the notes and categorized them into five different categories which we described briefly below with examples.

\subsection{Findings}
We randomly sampled 100 cross-platform fork candidates and manually analyzed the corresponding information. We classified the forks in five categories based on their activity history. 

\subsubsection{ Forks created using mirroring feature}
During our manual analysis, we find most of the forks are created using the mirror feature of GitLab. The mirror feature in GitLab crawls the original GitHub repository every five minutes to keep the forked repository in sync. For example, when we investigated cross-platform forks for \emph{Bitcoin} \citep{bitcoin}, we find a project \emph{SeppPenner/bitcoin}~\citep{sepp} in GitLab where the description section mentions the project is a mirror of the original \emph{Bitcoin} project. Also, it mentions that the project was updated 17 minutes ago which means GitLab was crawling it with some time interval. Later, we found from GitLab documentation\footnote{https://docs.gitlab.com/ee/user/project/repository/repository\_mirroring.html} that their mirror can be used to push or pull from or to a remote repository which is automatically updated in 5-minute intervals. This feature is a key for a significant number of cross-platform forks.

\subsubsection{ Forks owner are also a contributor of the original project}

Some cross-platform forks are owned by one of the contributors of the original repository and the last commit in the cross-platform fork belongs to the origin repository, whose committer is the fork owner. We verified that this commit belongs to the same person by comparing the usernames in GitHub and GitLab. For example, we found a project \emph{vectorci/tensorflow}~\citep{tensorflow2} which is a cross-platform fork of \emph{tensorflow/tensorflow}~\citep{tensorflow} on GitHub, where the owner is \emph{vectori}, who is also the committer of the last commit in GitLab which belongs to the original project. We found the account under username \emph{vectori} does not exist anymore. We suspect that the developer might want to leave Github and make GitLab as their new version control hosting platform. Therefore, to preserve their contribution to the main \emph{TensorFlow} repository, the developer forked it to GitLab.

\subsubsection{Forks which title renamed after forking}
Another kind of cross-platform forks has different name compared to the original project. During our manual analysis, we came across these kind of forks. For example, we found a repository on GitLab named \emph{onyx-gameboost}~\citep{onyx} in description section of which says \emph{``Onyx GameBoost Dev Repo''}. From the SWHGD we detected this repository as a cross-platform fork of the \emph{TensorFlow} repository, although all the commits in the GitLab copy is originated from the original \emph{TensorFlow} repository. From observing this type of cross-platform forks, we can possibly conclude that some forks are created and renamed for some purpose. Later, developers did not go through their intentions.

\subsubsection{Forks intended for an individual copy}
We found almost 70\% of the cross-platform forks are just merely for having a copy in different social coding platform accounts of the developers. As different social coding platforms provide different facilities, developers prefer to have copies of their own in different platforms for the projects they are interested in. All the cross-platform forks we found are one to four years older and they are inactive. Therefore, we can safely conclude that they are for keeping an individual copy. 

\subsubsection{Forks intended to continue development in another social coding platform}
From our analysis, we managed to retrieve one worth mentioning example. We found a repository named \emph{mg-vim}~\citep{mgvim} on GitLab which is a cross-platform fork of original \emph{Vim}~\citep{vim} repository. We found a commit which is absent in the original repository. From investigating the commit, we detected that it was due to adding CI/CD support provided by GitLab for the repository. Additionally, we studied about CI/CD support provided by different social coding platforms\footnote{https://usersnap.com/blog/gitlab-github/}. GitLab offers free CI/CD support developed by them. On the contrary, GitHub allows third-party CI/CD tools to be integrated on their platform but does not provide any support themselves when the fork created, although recently Github Actions started supporting free CI/CD for public repositories. Further investigation and interviews with developers can reveal more concrete reasons over cross-platform forks.

\section{Threats to Validity}
One might question the generality of our findings since we only used 10 projects from GitHub. In order to at least partially mitigate this issue, we carefully chose projects of diverse varieties and thus our findings could be generalizable to some extent. Of course analyzing more data might provide more insights for forking. However, by manually investigating the sampled projects, we reach the saturation and the findings already seems interesting and important. 


\section{Conclusion and Future Work}

To sum up, we conducted a mixed-method study (quantitative and qualitative) to identify cross-platform forks using the SWHGD. For our experiment, we started with 10 popular GitHub repositories and then detected their forks among all the platforms. Based on our qualitative analysis, we reported five types of cross-platform fork scenarios with concrete examples. In future, we plan to conduct a large scale analysis along with interviews with the developers to understand the pros and cons of different platforms, help stakeholders to make deliberate decisions on choosing code hosting platforms, and help platform providers to better facilitate social coding activities. Another research direction could be designing methods to track activities among cross-platform forks and generate a larger overview for cross-platform forks as a whole along with the idea proposed by Zhou et al.~\citep{zhou2018identifying}.

\textbf{Acknowledgments:} This work is supported in part by the Canada First Research Excellence Fund (CFREF) under the Global Institute for Water Security (GIWS).
\bibliographystyle{ACM-Reference-Format}
\bibliography{reference}


\begin{thebibliography}{19}


\ifx \showCODEN    \undefined \def \showCODEN     #1{\unskip}     \fi
\ifx \showDOI      \undefined \def \showDOI       #1{#1}\fi
\ifx \showISBNx    \undefined \def \showISBNx     #1{\unskip}     \fi
\ifx \showISBNxiii \undefined \def \showISBNxiii  #1{\unskip}     \fi
\ifx \showISSN     \undefined \def \showISSN      #1{\unskip}     \fi
\ifx \showLCCN     \undefined \def \showLCCN      #1{\unskip}     \fi
\ifx \shownote     \undefined \def \shownote      #1{#1}          \fi
\ifx \showarticletitle \undefined \def \showarticletitle #1{#1}   \fi
\ifx \showURL      \undefined \def \showURL       {\relax}        \fi
\providecommand\bibfield[2]{#2}
\providecommand\bibinfo[2]{#2}
\providecommand\natexlab[1]{#1}
\providecommand\showeprint[2][]{arXiv:#2}

\bibitem[\protect\citeauthoryear{??}{git}{2020}]%
        {githubDitched}
 \bibinfo{year}{2020}\natexlab{}.
\newblock \bibinfo{booktitle}{\emph{13,000 Projects Ditched GitHub for GitLab
  Monday Morning}}.
\newblock
\urldef\tempurl%
\url{https://www.vice.com/en\_us/article/ywen8x/13000-projects-ditched-github-for-gitlab-monday-morning}
\showURL{%
\tempurl}


\bibitem[\protect\citeauthoryear{??}{bit}{2020}]%
        {bitcoin}
 \bibinfo{year}{2020}\natexlab{}.
\newblock \bibinfo{booktitle}{\emph{Bitcoin}}.
\newblock
\urldef\tempurl%
\url{https://github.com/bitcoin/bitcoin}
\showURL{%
\tempurl}


\bibitem[\protect\citeauthoryear{??}{cro}{2020a}]%
        {cross_fork_dataset}
 \bibinfo{year}{2020}\natexlab{a}.
\newblock \bibinfo{booktitle}{\emph{{Cross-platform forks dataset}}}.
\newblock
\urldef\tempurl%
\url{https://doi.org/10.5281/zenodo.3699031}
\showDOI{\tempurl}


\bibitem[\protect\citeauthoryear{??}{mgv}{2020}]%
        {mgvim}
 \bibinfo{year}{2020}\natexlab{}.
\newblock \bibinfo{booktitle}{\emph{mg-vim}}.
\newblock
\urldef\tempurl%
\url{https://gitlab.com/mg_pub_group1/mg-vim}
\showURL{%
\tempurl}


\bibitem[\protect\citeauthoryear{??}{ony}{2020}]%
        {onyx}
 \bibinfo{year}{2020}\natexlab{}.
\newblock \bibinfo{booktitle}{\emph{Onyx GameBoost Dev Repo}}.
\newblock
\urldef\tempurl%
\url{https://gitlab.com/onyxdevteam/onyx-gameboost}
\showURL{%
\tempurl}


\bibitem[\protect\citeauthoryear{??}{cro}{2020b}]%
        {cross_fork_software}
 \bibinfo{year}{2020}\natexlab{b}.
\newblock \bibinfo{booktitle}{\emph{{Replication Package.}}}
\newblock
\urldef\tempurl%
\url{https://doi.org/10.5281/zenodo.3699105}
\showDOI{\tempurl}


\bibitem[\protect\citeauthoryear{??}{sep}{2020}]%
        {sepp}
 \bibinfo{year}{2020}\natexlab{}.
\newblock \bibinfo{booktitle}{\emph{SeppPenner/bitcoin}}.
\newblock
\urldef\tempurl%
\url{https://gitlab.com/SeppPenner/bitcoin}
\showURL{%
\tempurl}


\bibitem[\protect\citeauthoryear{??}{sql}{2020}]%
        {sqlalchemy}
 \bibinfo{year}{2020}\natexlab{}.
\newblock \bibinfo{booktitle}{\emph{Sqlalchemy}}.
\newblock
\urldef\tempurl%
\url{https://github.com/sqlalchemy/sqlalchemy}
\showURL{%
\tempurl}


\bibitem[\protect\citeauthoryear{??}{ten}{2020a}]%
        {tensorflow}
 \bibinfo{year}{2020}\natexlab{a}.
\newblock \bibinfo{booktitle}{\emph{Tensorflow}}.
\newblock
\urldef\tempurl%
\url{https://github.com/tensorflow/tensorflow}
\showURL{%
\tempurl}


\bibitem[\protect\citeauthoryear{??}{ten}{2020b}]%
        {tensorflow2}
 \bibinfo{year}{2020}\natexlab{b}.
\newblock \bibinfo{booktitle}{\emph{vectorci/tensorflow}}.
\newblock
\urldef\tempurl%
\url{https://gitlab.com/vectorci/tensorflow}
\showURL{%
\tempurl}


\bibitem[\protect\citeauthoryear{??}{vim}{2020}]%
        {vim}
 \bibinfo{year}{2020}\natexlab{}.
\newblock \bibinfo{booktitle}{\emph{Vim}}.
\newblock
\urldef\tempurl%
\url{https://github.com/vim/vim}
\showURL{%
\tempurl}


\bibitem[\protect\citeauthoryear{Gousios}{Gousios}{2013}]%
        {Gousi13}
\bibfield{author}{\bibinfo{person}{Georgios Gousios}.}
  \bibinfo{year}{2013}\natexlab{}.
\newblock \showarticletitle{The GHTorrent dataset and tool suite}. In
  \bibinfo{booktitle}{\emph{Proceedings of the 10th Working Conference on
  Mining Software Repositories}} (San Francisco, CA, USA)
  \emph{(\bibinfo{series}{MSR '13})}. \bibinfo{publisher}{IEEE Press},
  \bibinfo{address}{Piscataway, NJ, USA}, \bibinfo{pages}{233--236}.
\newblock
\showISBNx{978-1-4673-2936-1}
\urldef\tempurl%
\url{http://dl.acm.org/citation.cfm?id=2487085.2487132}
\showURL{%
\tempurl}


\bibitem[\protect\citeauthoryear{Gousios, Pinzger, and Deursen}{Gousios
  et~al\mbox{.}}{2014}]%
        {gousios2014exploratory}
\bibfield{author}{\bibinfo{person}{Georgios Gousios}, \bibinfo{person}{Martin
  Pinzger}, {and} \bibinfo{person}{Arie~van Deursen}.}
  \bibinfo{year}{2014}\natexlab{}.
\newblock \showarticletitle{An exploratory study of the pull-based software
  development model}. In \bibinfo{booktitle}{\emph{Proceedings of the 36th
  International Conference on Software Engineering}}. ACM,
  \bibinfo{pages}{345--355}.
\newblock


\bibitem[\protect\citeauthoryear{Pietri, Spinellis, and Zacchiroli}{Pietri
  et~al\mbox{.}}{2019}]%
        {pietri2019software}
\bibfield{author}{\bibinfo{person}{Antoine Pietri}, \bibinfo{person}{Diomidis
  Spinellis}, {and} \bibinfo{person}{Stefano Zacchiroli}.}
  \bibinfo{year}{2019}\natexlab{}.
\newblock \showarticletitle{The software heritage graph dataset: public
  software development under one roof}. In
  \bibinfo{booktitle}{\emph{Proceedings of the 16th International Conference on
  Mining Software Repositories}}. IEEE Press, \bibinfo{pages}{138--142}.
\newblock


\bibitem[\protect\citeauthoryear{Pietri, Spinellis, and Zacchiroli}{Pietri
  et~al\mbox{.}}{2020}]%
        {MSR20DC}
\bibfield{author}{\bibinfo{person}{Antoine Pietri}, \bibinfo{person}{Diomidis
  Spinellis}, {and} \bibinfo{person}{Stefano Zacchiroli}.}
  \bibinfo{year}{2020}\natexlab{}.
\newblock \showarticletitle{The {Software Heritage Graph Dataset}: Large-scale
  Analysis of Public Software Development History}. In
  \bibinfo{booktitle}{\emph{MSR 2020: The 17th International Conference on
  Mining Software Repositories}}. \bibinfo{publisher}{IEEE}.
\newblock


\bibitem[\protect\citeauthoryear{Shurui, Bogdan, and Christian}{Shurui
  et~al\mbox{.}}{2019}]%
        {zhou2020forkChange}
\bibfield{author}{\bibinfo{person}{Zhou Shurui}, \bibinfo{person}{Vasilescu
  Bogdan}, {and} \bibinfo{person}{Kästner Christian}.}
  \bibinfo{year}{2019}\natexlab{}.
\newblock \showarticletitle{How Has Forking Changed in the Last 20 Years? A
  Study of Hard Forks on GitHub}. In \bibinfo{booktitle}{\emph{ICSE 2020,
  23–29 May, 2020, Seoul, South Korea}}. IEEE, \bibinfo{pages}{230--241}.
\newblock


\bibitem[\protect\citeauthoryear{Steinmacher, Pinto, Wiese, and
  Gerosa}{Steinmacher et~al\mbox{.}}{2018}]%
        {steinmacher2018almost}
\bibfield{author}{\bibinfo{person}{Igor Steinmacher}, \bibinfo{person}{Gustavo
  Pinto}, \bibinfo{person}{Igor~Scaliante Wiese}, {and}
  \bibinfo{person}{Marco~Aur{\'e}lio Gerosa}.} \bibinfo{year}{2018}\natexlab{}.
\newblock \showarticletitle{Almost there: A study on quasi-contributors in
  open-source software projects}. In \bibinfo{booktitle}{\emph{2018 IEEE/ACM
  40th International Conference on Software Engineering (ICSE)}}. IEEE,
  \bibinfo{pages}{256--266}.
\newblock


\bibitem[\protect\citeauthoryear{Zhou, Stanciulescu, Le{\ss}enich, Xiong,
  Wasowski, and K{\"a}stner}{Zhou et~al\mbox{.}}{2018}]%
        {zhou2018identifying}
\bibfield{author}{\bibinfo{person}{Shurui Zhou}, \bibinfo{person}{Stefan
  Stanciulescu}, \bibinfo{person}{Olaf Le{\ss}enich}, \bibinfo{person}{Yingfei
  Xiong}, \bibinfo{person}{Andrzej Wasowski}, {and} \bibinfo{person}{Christian
  K{\"a}stner}.} \bibinfo{year}{2018}\natexlab{}.
\newblock \showarticletitle{Identifying features in forks}. In
  \bibinfo{booktitle}{\emph{2018 IEEE/ACM 40th International Conference on
  Software Engineering (ICSE)}}. IEEE, \bibinfo{pages}{105--116}.
\newblock


\bibitem[\protect\citeauthoryear{Zhou, Vasilescu, and K{\"a}stner}{Zhou
  et~al\mbox{.}}{2019}]%
        {zhou2019fork}
\bibfield{author}{\bibinfo{person}{Shurui Zhou}, \bibinfo{person}{Bogdan
  Vasilescu}, {and} \bibinfo{person}{Christian K{\"a}stner}.}
  \bibinfo{year}{2019}\natexlab{}.
\newblock \showarticletitle{What the fork: a study of inefficient and efficient
  forking practices in social coding}. In \bibinfo{booktitle}{\emph{Proceedings
  of the 2019 27th ACM Joint Meeting on European Software Engineering
  Conference and Symposium on the Foundations of Software Engineering}}. ACM,
  \bibinfo{pages}{350--361}.
\newblock


\end{thebibliography}

\end{document}